\documentclass[12pt,twoside]{article}
\usepackage{fleqn,espcrc1}

\usepackage{graphicx}
\usepackage[figuresright]{rotating}

\newcommand{\AmS}{{\protect\the\textfont2
  A\kern-.1667em\lower.5ex\hbox{M}\kern-.125emS}}

\title{The NNLO QCD analysis of the CCFR data for $xF_3$ : \\
$Q^2$-dependence of the parameters.}

\author{A.L. Kataev\address{Institute for Nuclear Research of the
        Academy of Sciences of Russia,\\
        117312 Moscow, Russia}
        \thanks{Supported in part by the Russian Foundation
        of Basic Research, Grant N 99-01-00091},
	G. Parente\address{Department of Particle Physics, University
	of Santiago de Compostela,\\
       15706 Santiago de Compostela, Spain}\thanks{Supported by
       CICYT (AEN96-1773) and Xunta de Galicia (Xuga-20602B98)}
        and
        A.V. Sidorov\address{Bogoliubov Laboratory of Theoretical Physics,
        Joint Institute for Nuclear Research, \\
        141980 Dubna, Russia}\thanks{Supported in part by the Russian Foundation
        of Basic Research, Grant N 99-01-0091.}}

\begin{document}

% typeset front matter
\maketitle

\begin{abstract}
We continue the systematic  next-to-next-to-leading order (NNLO)
QCD analysis of the CCFR data for the $xF_3$ SF.
The  NNLO results for the $Q^2$-dependence of the parameters,
which describe
the $x$-dependence  of this SF,
are presented.
\end{abstract}

\section{Introduction}

In the series of the subsequent papers\cite{KKPS1}-\cite{KPS2}
we concentrated ourselves on the fits of the CCFR data for
$xF_3$ SF \cite{CCFR} at the
LO, NLO and NNLO level
with twist-4 effects included through the
infrared-renormalon model of Refs.\cite{DW,Maul}. These fits
were done using the Jacobi polynomial-Mellin moments version
of the DGLAP method, developed and used in Refs.\cite{PS}-
\cite{Kriv}. In the process of these fits we have taken into account
the results
of calculations of  the NNLO corrections to the coefficient function
of $xF_3$ SF \cite{VZ}
and the available analytical expressions for the NNLO corrections
to the anomalous dimensions of the nonsinglet (NS) even moments with
$n=2,4,6,8,10$\cite{LRV}, supplemented with the given in Ref.\cite{KKPS1}
$n=3,5,7,9$ similar numbers, obtained using the smooth interpolation
procedure, which was previously proposed in Ref.\cite{PKK}.

The Jacobi polynomial method is based on the reconstruction
of the SF
from its moments $M_n^{NS}=\int_0^1 x^{n-1}F_3(x,Q^2)dx$
via the following equation:
\begin{equation}
xF_3^{N_{max}}=x^{\alpha}(1-x)^{\beta}\sum_{n=0}^{N_{max}}
\Theta_n^{\alpha,\beta}(x) \sum_{j=0}^{n}c_j^{(n)}(\alpha,\beta)
M_{j+2}^{NS}(Q^2)
%+\frac{h(x)}{Q^2}
\end{equation}
where $n=2,3,..$, $c_j^{(n)}(\alpha,\beta)$ is the combination of Euler
$\Gamma$-functions.
% and $h(x)$ is the twist-4 contribution.
For simplicity we present here the basic formula in the case
when target mass corrections  and higher-twist effects are neglected.
More details can be found in Ref.\cite{KPS2}, where the formulae for
all these
contributions  are given.

Making  the
fits with the free Jacobi polynomial parameters $\alpha,\beta$,
we found that their values
$\alpha\approx 0.7$, $\beta\approx 3$
correspond to  the minimum in the plane $(\alpha,\beta)$.

Using the standard model parameterization for the NS
$xF_3$ SF, we are now defining  the
corresponding moments
at the initial scale $Q_0^2$:
\begin{equation}
M_n^{NS}(Q_0^2)=\int_0^1 x^{n-2}
A(Q_0^2)x^{b(Q_0^2)}(1-x)^{c(Q_0^2)}(1+\gamma(Q_0^2) x)dx
\end{equation}
and evolve them to other  transfered momentum
using the solution of the renormalization-group  QCD equations for
the moments, which has the following form:
\begin{equation}
\frac{M_n^{NS}(Q^2)}{M_n^{NS}(Q_0^2)}=
\bigg(\frac{A_s(Q^2)}{A_s(Q_0^2)}\bigg)^{\frac{\gamma_{NS}^{(0)}}{2\beta_0}}
\frac{AD(n,Q^2)C^{(n)}(Q^2)}{AD(n,Q_0^2)C^{(n)}(Q_0^2)}
\end{equation}
where $A_s(Q^2)=\alpha_s(Q^2)/(4\pi)$ is the $\overline{MS}$-scheme
expansion parameter, $AD(n,Q^2)=1+p(n)A_s(Q^2)+q(n)A_s(Q^2)^2+...$
comes from the expansion of the anomalous dimensions term of
the solution of the corresponding renormalization-group equation,
$C^{(n)}(Q^2)=1+C^{(1)}(n)A_s(Q^2)+C^{(2)}(n)A_s(Q^2)^2+...$ is
the coefficient function of the $n$-th moment of the $xF_3$ SF.
The numerical values  of the coefficients $p(n)$, $q(n)$, $C^{(1)}(n)$
and $C^{(2)}(n)$, normalized to the case of $f=4$ numbers of active flavours,
are given in Ref.\cite{KPS1,KPS2}.
In our fits we will use well-known LO, NLO and NNLO
expressions of $\alpha_s$, written down as the
expansions in  the inverse powers
of
$ln(Q^2/\Lambda_{\overline{MS}}^{(4)~2})$-terms in the LO, NLO and NNLO.

\section{Presentation of the results}

\begin{table}[p]
\caption{
 The $Q_0^2$-dependence of the parameters of $xF_3$ model,
 extracted from the fits of CCFR'97 data with the cut $Q^2>
 5~ GeV^2$. The statistical errors are taken into account.  }
\label{table:param}
\newcommand{\m}{\hphantom{$-$}}
\newcommand{\cc}[1]{\multicolumn{1}{c}{#1}}
\renewcommand{\arraystretch}{1.2} % enlarge line spacing
\begin{tabular}{@{}ccccccc}
\hline
 $Q_0^2=5~GeV^2$& $\Lambda_{\overline{MS}}^{(4)}$ [MeV]&    A      &  b
& c
&
$\gamma$
& $\chi^2$/n.e.p.   \\
\hline
 LO    &266 $\pm$ 35& 5.13 $\pm$ 0.46&  0.72 $\pm$ 0.03&  3.87 $\pm$ 0.05&
1.42 $\pm$ 0.33&   113.2/86   \\
 NLO   &341 $\pm$ 30& 4.05 $\pm$ 0.36&  0.65 $\pm$ 0.03&  3.71 $\pm$ 0.06&
1.93 $\pm$ 0.36&    87.1/86   \\
 NNLO  &293 $\pm$ 29& 4.25 $\pm$ 0.38&  0.66 $\pm$ 0.03&  3.56 $\pm$
0.07&
1.33 $\pm$ 0.33&    78.4/86   \\
\hline
 $Q_0^2=8~GeV^2$& &&&&&   \\
 LO    &266 $\pm$ 35& 5.10 $\pm$ 0.15&  0.70 $\pm$ 0.01&  3.94 $\pm$ 0.04&  1.23 $\pm$ 0.12&   113.2/86   \\
 NLO   &340 $\pm$ 40& 4.38 $\pm$ 0.17&  0.66 $\pm$ 0.02&  3.81 $\pm$ 0.07&  1.47 $\pm$ 0.19&    87.4/86   \\
 NNLO  &312 $\pm$ 33& 4.42 $\pm$ 0.36&  0.66 $\pm$ 0.03&  3.68 $\pm$ 0.07&  1.13 $\pm$ 0.31&    76.5/86   \\
\hline
 $Q_0^2=10~GeV^2$& &&&&& \\
 LO   & 265 $\pm$ 34& 5.07 $\pm$ 0.15&  0.70 $\pm$ 0.01&  3.97 $\pm$ 0.04&  1.16 $\pm$ 0.12&   113.2/86   \\
 NLO  & 340 $\pm$ 35& 4.48 $\pm$ 0.15&  0.66 $\pm$ 0.02&  3.85 $\pm$ 0.04&  1.32 $\pm$ 0.10&    87.5/86   \\
 NNLO & 318 $\pm$ 33& 4.50 $\pm$ 0.36&  0.65 $\pm$ 0.03&  3.73 $\pm$ 0.07&  1.05 $\pm$ 0.31&    76.3/86   \\
\hline
 $Q_0^2=15~GeV^2$&  &&&&& \\
 LO   & 265 $\pm$ 35& 5.02 $\pm$ 0.16&  0.68 $\pm$ 0.01&  4.01 $\pm$ 0.04&  1.04 $\pm$ 0.11&   113.1/86   \\
 NLO  & 339 $\pm$ 37& 4.61 $\pm$ 0.42&  0.65 $\pm$ 0.03&  3.92 $\pm$ 0.11&  1.08 $\pm$ 0.42&    87.6/86   \\
 NNLO & 324 $\pm$ 34& 4.60 $\pm$ 0.34&  0.65 $\pm$ 0.03&  3.83 $\pm$ 0.07&  0.92 $\pm$ 0.29&    76.6/86   \\
\hline
 $Q_0^2=20~GeV^2$&  &&&&& \\
 LO   & 264 $\pm$ 35& 4.98 $\pm$ 0.15&  0.68 $\pm$ 0.01&  4.05 $\pm$ 0.04&  0.96 $\pm$ 0.12&   113.1/86   \\
 NLO  & 339 $\pm$ 36& 4.67 $\pm$ 0.16&  0.65 $\pm$ 0.02&  3.96 $\pm$ 0.05&  0.95 $\pm$ 0.13&    87.6/76   \\
 NNLO & 326 $\pm$ 35& 4.70 $\pm$ 0.34&  0.65 $\pm$ 0.03&  3.88 $\pm$0.08&  0.80 $\pm$ 0.30&    77.0/86   \\
\hline
 $Q_0^2=30~GeV^2$&  &&&&& \\
 LO    &264 $\pm$ 37& 4.92 $\pm$ 0.17&  0.67 $\pm$ 0.02&  4.09 $\pm$ 0.05&  0.86 $\pm$ 0.09&   113.0/86   \\
 NLO   &338 $\pm$ 37& 4.72 $\pm$ 0.43&  0.64 $\pm$ 0.03&  4.01 $\pm$ 0.08&  0.80 $\pm$ 0.35&    87.5/86   \\
 NNLO  &327 $\pm$ 35& 4.78 $\pm$ 0.32&  0.64 $\pm$ 0.03&  3.96 $\pm$ 0.08&  0.67 $\pm$ 0.27&    77.8/86   \\
\hline
 $Q_0^2=50~GeV^2$&  &&&&& \\
 LO    &264 $\pm$ 35& 4.84 $\pm$ 0.15&  0.65$\pm$ 0.01&  4.13 $\pm$ 0.04&  0.75 $\pm$ 0.12&   112.9/86   \\
 NLO   &337 $\pm$ 34& 4.74 $\pm$ 0.13&  0.64$\pm$ 0.01&  4.06 $\pm$ 0.06&  0.64 $\pm$ 0.14&    87.5/86   \\
 NNLO  &326 $\pm$ 36& 4.85 $\pm$ 0.31&  0.64$\pm$ 0.02&  4.03 $\pm$ 0.09&  0.53 $\pm$ 0.27&    78.8/86   \\
\hline
$Q_0^2=100~GeV^2$&  &&&&& \\
 LO   & 263 $\pm$ 36& 4.73 $\pm$ 0.23&  0.64$\pm$ 0.02&  4.19 $\pm$ 0.09&  0.62 $\pm$ 0.24&   112.6/86   \\
 NLO  & 337 $\pm$ 37& 4.73 $\pm$ 0.15&  0.62$\pm$ 0.02&  4.12 $\pm$ 0.06&  0.46 $\pm$ 0.12&    87.3/86   \\
 NNLO & 325 $\pm$ 36& 4.91 $\pm$ 0.28&  0.63$\pm$ 0.02&  4.11 $\pm$ 0.10&
0.36 $\pm$ 0.25&    80.0/86   \\
\hline
\end{tabular}\\[2pt]
\end{table}

Here we complete our previous  analysis of the CCFR'97 data of
Refs.\cite{KKPS2,KPS1,KPS2} 
and concentrate our attention on the extraction of the $Q^2$
dependence of the paremeters
of the model of Eq.(2) for $xF_3$ SF (which, of course, are related 
to the parton distributions parameters).   
In Table 1  we present the results of our
fits of the CCFR $xF_3$ data with the cut $Q^2>5~GeV^2$
for different fixed falues of $Q_0^2$ with TMC taken  into account,
but twist-4 contributions neglected.
Several comments are in order:
%\begin{itemize} 

1) For the $x$-shape parameters the shift between LO and NLO values
is larger than between NLO and NNLO results (exept parameter $\gamma$ ).
This fact indicates the supression of the uncertainties 
of the values of the parameters as the result    of taking into
account  higher order perturbative QCD effects.
%\item 

2) The numerical values of the  parameters, obtained 
in the fits at the NNLO, 
are
in a good agreement with those obtained in Ref. \cite{ST}
where HT corrections were
taken into account and the nuclear effects parameterized
by deuteron model approximation of $xF_3^{Fe}$ \cite{STnucl}.
%\item 

3) The effect of the increase of  the values  of the  parameter $c$
and
the simultaneous
decrease of the values of the parameter $b$,
which is observed while
incresing the value of $Q_0^2$-scale,  is in a qualitative agreement
with the  theoretical predictions of the works of Ref.\cite{Korch} and
Ref.\cite{Mona}.
%\item 

4) The value of the parameter $b$, which governs the
low-$x$ behaviour of the $xF_3$ SF, is in agreement with the recent 
theoretical calculation of the low-$x$ behaviour of the  
NS structure functions \cite{EGT}.
%\item

5) In the case of $Q_0^2=5~GeV^2$ the NLO results for the 
parameters $b$ and $c$ are in agreement with the 
outcomes of the NLO fits of the CCFR'97 data, which were 
made in Ref.\cite{AK} with the help of the standard realization 
of the DGLAP equation. Note, however, that the value of 
parameter $\gamma$, obtained in Ref.\cite{AK}, turned out 
to be comparable with zero. It is of interest to trace 
the origin of this only
difference with
the presented in Table 1 results. 
%\item

6) One can see that the values of $\Lambda_{\overline{MS}}^{(4)}$, which 
come from the 
LO and
NLO fits, are
rather stable to the  choice of the initial scale $Q_0^2$.
However, the results of the NNLO fits are more  sensitive
to its variation. Indeed, despite the fact that 
for smaller values of $Q_0^2$ the difference between 
NLO and NNLO values of the parameters $A,~b,~c,~\gamma$ 
are not large, the NNLO values of $\Lambda_{\overline{MS}}^{(4)}$
are considerably smaller, than the NLO ones. This effect 
is decreasing
while increasing $Q_0^2$ from
$5~GeV^2$ up to $20~GeV^2$. For $Q^2\geq 20~GeV^2$ 
the NNLO values of $\Lambda_{\overline{MS}}^{(4)}$ are becoming
rather
stable. 
%\item

7) We think that this effects is related to rather
peculiar behavior of the NNLO perturbative QCD expansion
of $n=2$ moment. Indeed, taking into account the exact
numerical values of the coefficients $p(2)$, $q(2)$, $C^{(1)}(2)$
and $C^{(2)}(2)$, given in Refs.\cite{KPS1,KPS2}, we find that the
perturbative behavior of the $n=2$ moment
is determined by the following series:
\begin{equation}
 AD(2,Q^2)C^{(2)}(Q^2)=1-0.132A_s(Q^2) -46.16\bigg(A_s(Q^2)\bigg)^2+...
\end{equation}
where the value of the relatively large
$A_s^2$ coefficient mainly comes from the  
NNLO term  of the coefficient function of $n=2$ moment.
Thus we conclude, that it is safer 
to start the QCD evolution from the
scale $Q_0^2=20~GeV^2$, where the numerical value
of the $A_s^2$ contribution in Eq.(4) is smaller, and
therefore the complicated asymptotic structure
of the perturbative expansion of $n=2$ moment is not
yet manifesting itself. Note, that the choice of this initial
scale is also empirically supported by the fact that
it is lying in the mid of the kinematic region of the CCFR
data.
%\item 

8) It is interesting to note, that the nonstandard  
structure of the 
perturbative QCD expression for Eq.(4) is manifesting itself 
not only in the case of $f=4$  number of flavours. Indeed, 
in the cases of $f=3$ and $f=5$  
the 
expressions for 
the NNLO approximations of the corresponding solutions 
of the renormalization-group equations read
\begin{eqnarray}
AD(2,Q^2)C^{(2)}(Q^2)&=&1-0.271 A_s(Q^2)-43.51\bigg( A_s(Q^2)\bigg)^2
+....\\ \nonumber
AD(2,Q^2)C^{(2)}(Q^2)&=&1+0.126 A_s(Q^2)-48.88\bigg(A_s(Q^2)\bigg)^2
+....
\end{eqnarray}
%\end{itemize}
Thus, we conclude, that the unnatural behaviour of  the 
order $O(A_s^2)$ approximation of the renormalization-group 
improved Mellin moment cannot be avoided after changing the number 
of active flavours.  
%\end{itemize}

In Table 2 we present the results of our  fits of the CCFR'97 data
including target mass corrections   both without
twist-4 corrections and with twist-4 contributions,
fixed through the IRR model of Ref.\cite{DW} as 
\begin{equation}
M_n^{IRR}=\tilde{C}(n)M_n^{NS}(Q^2)\frac{A_2^{'}}{Q^2}
\end{equation}
where $\tilde{C}(n)=-n-4+2/(n+1)+4/((n+2)+4S_1(n)$ 
($S_1(n)=\sum_{j=1}^{n}1/j$) and $A_2^{'}$ is the free parameter.

\begin{table}
\caption{The results of the fits of CCFR'97 data 
with the cut $Q^2>5~GeV^2$, obtained in the case  
$Q_0^2=20~GeV^2$. N$^3$LO means the application 
of [0/2] expanded Pad\'e approximants. }
\label{table:res}
\newcommand{\cc}[1]{\multicolumn{0.8}{c}{#1}}
\renewcommand{\arraystretch}{1}
\begin{tabular}{@{}ccccccc}
\hline
  $\Lambda_{\overline{MS}}^{(4)}$ [MeV] &A & b& c & 
$\gamma$ & $A_2^{'} [GeV^2]$ & $\chi^2$/n.e.p. \\
\hline
LO~~~~ $264\pm 36$ & $4.98\pm 0.23$  & $0.68\pm0.02$ & $4.05\pm0.05$ 
& $0.96\pm 0.18$ & -----& 113.1/86 \\
~~~~~~~ $433\pm 51$ & $4.69\pm 0.13$  & $0.64\pm 0.01$ & $4.03\pm0.04$
& $1.16\pm0.12$  & $-0.33\pm 0.12$ & 83.1/86 \\ 
\hline
NLO~~~ $339\pm 35$ & $4.67\pm 0.11$ & $0.65\pm 0.01$ & $3.96\pm 0.04$
& $0.95\pm 0.09$ & -----& 87.6/86 \\
~~~~~~~  $369\pm37$ & $4.62\pm 0.16$ & $0.64\pm 0.01$ & $3.95\pm 0.05$
& $0.98\pm 0.17$ & $-0.12\pm0.06$ & 82.3/86 \\
\hline
NNLO~~ $326\pm 35$ & $4.70\pm0.34$ & $0.65\pm 0.03$ & $3.88\pm 0.08$
& $0.80\pm 0.28$ & ----- & 77.0/86 \\
~~~~~~~~ $327\pm 35$ & $4.70\pm 0.34$ & $0.65\pm 0.03$ & $3.88\pm 0.08$& 
$0.80\pm 0.29$ &
$-0.01\pm 0.05$ &76.9/86 \\
\hline 
N$^3$LO~~ $335\pm 37$& $4.77\pm0.34$& 
$0.65\pm  0.03$ & $3.85\pm0.08$& $0.71\pm 0.28$ &
----- & 77.9/86 \\
(Pade) $340\pm  37$ & $4.78\pm 0.34$& $0.65\pm 0.03$ & 
$3.85\pm 0.08$ & $0.71\pm 0.28$ & $-0.04\pm 0.05$ & 77.2/86 \\
\hline   
\end{tabular}\\[1pt]
\end{table}
%\\[-1.5cm]
The theoretical uncertainties of the NNLO results are 
probed by using the [0/2] Pade expanded approximants 
in Eq.(3) and using the N$^3$LO expression for 
the expansion parameter $A_s$, which depends 
from the calculated recently four-loop coefficient 
of the QCD $\beta$-function \cite{RVL}.

The problem of comparison of the results of the 
fits with the IRR-model estimates and 
the effect of decreasing the value of its parameter 
$A_2^{'}$ at the NNLO was analysed in detail 
in Ref.\cite{KKPS2} and Ref.\cite{KPS2} in particular.
Here we would like to draw the attention to 
the fact, that besides the NLO value of 
$\Lambda_{\overline{MS}}^{(4)}$ (and thus $\alpha_s$) 
is sensitive to the twist-4 contributions, modeled 
with the help of the IRR approach, the parameters
of the $x$-dependence of the $xF_3$ SF (and 
thus parton distribution parameters) are more 
sensitive to the choice of $Q_0^2$-scale, than to 
the twist-4 effects (see Tables 1,2 above and Table 5
in Ref.\cite{KPS2}). Taking into account 
this effects might allow to fix theoretical uncertainties 
of the parton distributions parameters. The importance 
of this theoretical problem was emphasized in Ref.\cite{Forte}. 

To conclude, let us present the NLO and NNLO values of $\alpha_s(M_Z)$,
obtained in Ref.\cite{KPS2} from the results of the fits 
of CCFR'97 data for $xF_3$ with twist-4 terms modeled 
through the IRR approach (see Table 2) and using the 
$\overline{MS}$-matching condition 
in different orders of perturbation theory \cite{BW,LRVM,ChKS}:
\begin{eqnarray}
NLO~~~\alpha_s(M_Z)&=&0.120 \pm 0.003 (stat) \pm 0.005 (syst) 
\pm 0.004(theor) \\ \nonumber 
NNLO~~\alpha_s(M_Z)&=&0.118 \pm 0.003 (stat) \pm 0.005 (syst)
\pm 0.003 (theor)
\end{eqnarray}
The systematical uncertainty in these results are 
determined by using  the 
original CCFR considerations and the theoretical errors
are fixed using the results of [0/2] Pade approximations  fits 
and the proposals of Ref.\cite{ShM,BV} to estimate  
the ambiguities due to smooth transition to the world 
with $f=5$ numbers of flavours. 
It should be noted that the NLO result is in agreement 
with the one, obtained in Ref.\cite{AK} using the DGLAP 
equation and taking into consideration the correlations of 
statistical and systematical uncertainties.

{\bf Acknowledgements}
We wish to thank J. Ch\'yla, Yu.L. Dokshitzer and P. Nason for 
stimulating useful 
questions to our work of Ref.\cite{KPS2}. 
One of us (ALK) wishes to thank 
INFN and Organizing Committee of Nucleon99 Workshop
for hospitality in Frascati and his colleagues 
from Genova Section of INFN for hospitality in Genova 
during the completing of the work on this contribution.

\end{document}